\documentclass[12pt]{article}

\usepackage{epsfig}
\usepackage{amssymb}
\usepackage{graphics}
\usepackage{feynmf}
\usepackage[vcentermath,enableskew]{youngtab}
\let\a=\alpha      \let\g=\gamma   \let\d=\delta
\let\e=\epsilon \let\z=\zeta        
      \let\l=\lambda  \let\m=\mu
\let\n=\nu

\def\sp{\;\;,\;\;}

\newcommand{\be}{\begin{equation}}
\newcommand{\ee}{\end{equation}}
\newcommand{\bea}{\begin{eqnarray}}
\newcommand{\eea}{\end{eqnarray}}
\newcommand{\ba}{\begin{array}}
\newcommand{\ea}{\end{array}}
\def\nn{\nonumber}

\begin{document}
\begin{titlepage}
\rightline{hep-th/0701114}
\rightline{ROM2F/2007/01}
\vskip 2cm
\centerline{{\large\bf Anomalous U(1)$'$s, Chern-Simons couplings
and the Standard Model}}
\vskip 0.3cm
\vskip 1cm
\centerline{
P. Anastasopoulos \footnote{Pascal.Anastasopoulos@roma2.infn.it}}
\vskip 1cm
\smallskip
\centerline{Dip. di Fisica \& Sez. I.N.F.N. \ Universit{\`a} di
Roma \ ``Tor Vergata'',}
\centerline{Via della Ricerca
Scientifica, 1 - 00133 \ Roma, \ ITALY}
\vskip 1cm
\begin{abstract}
This proceeding is based on hep-th/0605225 and it shows that the
most general anomaly related effective action contains
St\"uckelberg, axionic and Chern-Simons-like couplings. Such
couplings are generically non-trivial in orientifold string vacua.
A similar analysis in quantum field theories provides similar
couplings. These Chern-Simons couplings generate new signals which
might be visible at LHC.
\end{abstract}
\end{titlepage}

%======================================================================
%\maketitle
\section{Introduction}
%======================================================================

Recently, many attempts have been made with partial success, in
order to embed the Standard Model (SM) in open string theory
\cite{Aldazabal:2000dg}-\cite{Anastasopoulos:2006da}. In such a
context the Standard Model particles are open string states
attached on (different) stacks of D-branes.
$N$ coincident D-branes typically generate a unitary group which
provides an anomalous U(1) under the usual decomposition U$(N)\to$
SU($N$)$\times$ U(1).

Such U(1)$'$s have generically 4d anomalies. The anomalies are
cancelled via the Green-Schwarz mechanism \cite{Green:sg,
Sagnotti:1992qw, Ibanez:1998qp, Bianchi:2000de} where a scalar
axionic field (zero-form, or its dual two-form) is responsible for
the anomaly cancellation. This mechanism gives a mass to the
anomalous U(1)$'$s and breaks the associated gauge symmetry. These
masses are typically of order of the string scale but in open
string theory they can be also much lighter \cite{Scrucca:2002is,
Antoniadis:2002cs}.
If the string scale is around a few TeV, observation of such
anomalous U$(1)$ gauge bosons becomes a realistic possibility
\cite{Kiritsis:2002aj, Ghilencea:2002da}.

As it has been shown in \cite{Antoniadis:2002cs, pascal}, we can
compute the general mass formuli of the anomalous U$(1)$$'$s in
supersymmetric and non-supersymmetric models by evaluating the
ultraviolet tadpole of the one-loop open string diagram with the
insertion of two gauge bosons on different boundaries.
It turns out that U$(1)$ gauge fields that are free of
four-dimensional anomalies can still be massive due to the
presence of mass-generating higher-dimensional anomalies
\cite{Ibanez:2001nd, Scrucca:2002is, Antoniadis:2002cs, pascal} .

However, Green-Schwarz mechanism is not enough to cancel all the
anomalies. Mixed abelian anomalies between anomalous and
non-anomalous factors need generalized Chern-Simons terms to be
cancelled. Here, we will stress the role of these terms by using a
toy-model. A more detailed study can be found in
\cite{Anastasopoulos:2006cz}.

All the above have very interesting phenomenological consequences
because if these anomalous U(1) masses are in the TeV range, these
fields behave like $Z'$ gauge bosons widely studied in the
phenomenological literature
\cite{Kiritsis:2002aj}-\cite{Anastasopoulos:2006cz}.
However, non trivial generalized Chern-Simons terms are needed to
cancel all the mixed anomalies between the non anomalous
(hypercharge) and the anomalous gauge bosons, generating new
signals that distinguish such models from other $Z'$ models. If
the string scale is of order of a few TeV, such signals may be
visible at LHC \cite{d-effective}.

\section{Generalized Chern-Simons: A toy model}\label{anomalies}
%======================================================================

In this section, we show that the Green-Schwarz mechanism is not
enough to cancel all the anomalies in models with various
U(1)$'$s. In particular, generalized Chern-Simons terms are
necessary to cancel mixed anomalies between anomalous and
non-anomalous U(1)$'$s \cite{Anastasopoulos:2006cz}.

We will concentrate on a toy-model which consists of a chiral
gauge theory with only two U(1)$'$s \footnote{Generalizations can
be found in \cite{Anastasopoulos:2006cz}}, one anomalous with
gauge field $A_{\mu}$, field strength $F^A_{\m\n}$ and charge
operator $Q_A$, and one non-anomalous with gauge field $Y_{\mu}$,
field strength $F^Y_{\m\n}$ and charge operator $Q_Y$.
%
%$G^a_{\m\n}$ denote other non-abelian field strengths.
%
By definition, the non-anomalous gauge boson obeys the conditions
\footnote{Where the trace runs onto all left-right fermions,
denoted by $f$: $Tr[Q^iQ^jQ^k]=\sum_f Q^i_fQ^j_fQ^k_f$.}:
$Tr[Q_Y]=Tr[Q_Y^3]=0$ but in general, other traces are non zero:
\bea Tr[Q_A Q_Y^2]=c_1\sp Tr[Q_A^2Q_Y]=c_2\sp Tr[Q_A^3]=c_3
~.~~~~\label{A}\eea
The above traces imply the following anomalous transformations of
the one-loop effective action. Under
\bea A_{\mu}\to A_{\mu}+\partial_{\mu}\epsilon \sp Y_{\mu}\to
Y_{\mu}+\partial_{\mu}\zeta\eea
the action transforms as:
\bea \delta S_{\rm 1-loop}&=&\int
d^4x~\bigg\{\epsilon\left[{c_3\over 3}F^A\wedge F^A +c_2~F^A\wedge
F^Y+c_1 ~F^Y\wedge F^Y
%+\xi~ Tr[G\wedge G]
\right] \nn\\
&& ~~~~~~~~ +\zeta\left[c_2~F^A\wedge F^A+c_1~F^A\wedge
F^Y\right]\bigg\}\eea

Following Green and Schwarz, we add the classical action:
\bea S_{\rm axion}&=&\int d^4 x~\Big\{-{1\over
4g_Y^2}(F^Y)^2-{1\over 4g_A^2}(F^A )^2+ (\partial_{\mu} \a + M
A_{\mu})^2\nn\\
&&~~~~\quad ~~~~~~+ \a \left(d_3~F^A\wedge
F^A+d_2~F^A\wedge F^Y+d_1~F^Y\wedge F^Y%+d_0~ trG\wedge G
\right)\Big\}\eea
where $d_1,~d_2,~d_3,~M$ are constants. The axion $\a$ transforms
as $\a\to \a - M \epsilon$ and we are assuming that $\a$ does not
shift under non-anomalous gauge transformations parameterized by
$\zeta$. Although $Q_A$ and $Q_Y$ mix in a certain sense ($Tr[Q_A
Q_Y] \neq 0$), we will confirm that $\a$ does not couple \'a la
St\"uckelberg to $Y_\mu$.

It is obvious that the axionic transformation does not cancel all
the anomalies. It is necessary to add non $\z$-invariant
$generalized$ $Chern-Simons$ terms:
\bea S_{\rm GCS}&=&\int Y\wedge A\wedge \Big\{d_4~F^{A} - d_5
F^{Y} \Big\} \eea
where all $d_i$ are constants. The gauge variation of the
classical action:
\bea \delta S_{\rm axion}+\delta S_{\rm GCS}&=&-\int
\epsilon~\Big\{ ~d_3~ F^A\wedge F^A+ (d_2-d_4)~F^A\wedge F^Y
+(d_1+d_5) ~F^Y\wedge F^Y %+ d_0~ tr(G\wedge G)
\Big\}\nn\\
%\delta S_{\rm GCS}&=&-\int
&&-\int \zeta~\Big\{d_4~F^A\wedge F^A -d_5~F^Y\wedge F^A\Big\}~.
\eea
Anomaly cancellation $\d S_{\rm 1-loop}+ \d S_{\rm axion}+\d
S_{\rm GCS}=0$ implies: $d_1=2c_1$, $d_2=2c_2$, $d_3=c_3/3$,
$d_4=c_2$, $d_5=-c_1$.
The presence of the generalized CS terms is due to the
non-vanishing mixed anomalies between the two U(1)$'$s (non
vanishing $c_{1}$, $c_2$).

We can generalize the previous example to the case of several
U(1)$'$s and axions. The anomaly-related terms in the effective
action are:
\bea {\cal S} & = & \int d^4x \Big[- \sum_i {1 \over 4 g_i^2}
F_{i,\m\n}  F_i^{\m\n} - {1 \over 2} \sum_I ( \partial_\m a^I +
M^I_i A_{\m}^i)^2
\ , \nonumber \\
&&\quad \quad \quad \quad + {1 \over 24 \pi^2} C_{ij}^I a^I
F^i\wedge F^j + {1 \over 24 \pi^2} E_{ij,k} A^i\wedge  A^j\wedge
F^k \ \Big], \label{i1} \eea
where $A_i$ are abelian gauge fields, $a^I$ are axions with
St\"uckelberg couplings which render massive (some of) the gauge
fields.

This action is gauge-variant under
%
%\be
$A^{i}\to A^i+d\e^i,~ a^I\to a^I-M^I_i\e^i .$
%\ee
%
This gauge-variance is tuned to cancel the anomalous variation of
the one-loop effective action due to the standard triangle graphs.
The contribution of the  triangle graphs is scheme dependent, (see
\cite{Anastasopoulos:2006cz} and references there for a detailed
exposition). In a natural scheme where the anomalous variation is
distributed democratically among the three vertices, the anomaly
cancellation conditions read:
\be t_{ijk}\ + \ E_{ijk}+E_{ikj} \ + \ M^I_i \ C^I_{jk} \ = \ 0 \
. \label{i2} \ee
Here $t_{ijk} = Tr (Q_i Q_j Q_k)$ are the standard anomaly traces
explained above and $Q_i$ is the charge generator associated to
$A_i$.

Notice that $t_{ijk}$ is a completely symmetric tensor
$\Yboxdim8pt\yng(3)$, $E_{ijk}$ is antisymmetric in the first two
indices $\Yboxdim8pt\yng(2,1)$ and $M^I_i ~C^I_{jk}$ is a sum of
two types of tensors: $\Yboxdim8pt\yng(3)$ and
$\Yboxdim8pt\yng(2,1)$. Therefore, there are special cases $eg$
model without fermions where the cancellation is achieved between
the axionic couplings and the GCS terms, or a model with a single
U(1) where the anomalies are cancelled by the axionic couplings
since $E_{ijk}$ is trivial. However, in general, all terms in
(\ref{i2}) are necessary.

\subsection{Generalized Chern-Simons in orientifold models}

Generically, anomalous and non-anomalous factors appear in
orientifolds \cite{Bianchi:1988fr}. Therefore, the presence of GCS
terms is necessary to cancel all the anomalies.

A direct string computation of these terms requires the evaluation
of the 1-loop open string amplitudes with the insertion of three
bosonic vertex operators (VO) in the odd spin structure $i.e.$ an
annulus with two VO$'$s and a single VO on the opposite
boundaries, and an annulus and a M\"obius strip with all three
VO$'$s on the same boundary. In the closed IR limit, the two last
diagrams cancel in any consistent theory due to tadpole
cancellation and the remaining non-planar cylinder diagram
contains the (antisymmetrized) Chan-Paton traces:
\be E_{ijk}=
{1\over 3}\sum_{\kappa} \eta_\kappa |\sqrt{N_\kappa}| \ tr
[\gamma_\kappa \l_{k} \l_{[j}] \ tr [\gamma_\kappa \l_{i]} ] \ .
\label{np3} \ee
Here $\kappa = 1 \cdots N-1$ denotes the different type of twisted
sectors propagating in the tree-level channel cylinder diagram,
whereas
\bea N_\kappa \ =\left\{
\begin{array}{ll}
\prod_{\Lambda=1}^3 (2 \sin [\pi \kappa v_\Lambda])^2 &~~~~
{\rm for \ D9-D9 \ and \ D5-D5 \ sectors,} \\
\\
(2 \sin [\pi \kappa v_3])^2 &~~~~ {\rm for \ D9-D5 \ sectors}
\end{array} \right.\label{Nk} \eea
denote the number of fixed points in the internal space and in the
third internal torus, respectively (we consider for simplicity D5
branes whose world-volume span the third internal torus $T^2_3$).
Also, $\eta_k$ depends on the sector and is given by
sign$(\prod_{\Lambda=1}^3 \sin [\pi k v_\Lambda])$ for all sectors
of D9-D9, D5-D5, D9-D5 where the orbifold action twists all tori,
$(-1)^{kv_i}$ for all sectors of D9-D5 where the orbifold action
leaves untwisted a perpendicular torus $T^2_i$ to the D5 brane
(all the above are ${\cal N}=1$ sectors), and zero for sectors of
D9-D9, D5-D5, D9-D5 where the orbifold action leaves untwists the
longitudinal torus $T^2_3$ to the D5-brane (which are ${\cal N}=2$
sectors).
Notice that particles and antiparticles contribute to the anomaly
with different signs as it should be.

Applying (\ref{np3}) in various 4d orientifold models, like $Z_6$
and $Z_6'$, we find that non-zero GCS terms are necessary to
cancel all the anomalies \cite{Anastasopoulos:2006cz}.

%\newpage

\subsection{Generalized Chern-Simons in effective filed theories}

An interesting question is wether in the anomaly sector of an EFT,
we can distinguish if the UV completion is stringy or an
UV-complete QFT.

We consider a consistent ({\it i.e.} anomaly-free) and
renormalizable gauge theory with spontaneously-broken gauge
symmetry via the Brout-Englert-Higgs mechanism. Through
appropriate Yukawa couplings, some large masses can be given to a
subset of the fermions. Absence of anomalies requires that $
\sum_{\rm light+Heavy} (Q_L^i Q_L^j Q_L^k - Q_R^i Q_R^j Q_R^k ) \
= \ 0$ where $Q_i$$'$s denote the charge operators of the various
U(1)$'$s. In general, the previous sum evaluated only for the
light fermions is different from zero, generating a superficially
anomalous EFT at a lower scale than the heavy fermion mass $M_H$.

It has been shown in \cite{Anastasopoulos:2006cz} that the
decoupling of heavy chiral fermions by large Yukawa couplings does
generate a generalized Green-Schwarz mechanism at low energy, with
axionic couplings cancelling anomalies of the light fermionic
spectrum in combination with generalized Chern-Simons terms which
play an important role in anomaly cancellation:
%
%======================================================
%=============        Diagrams        =================
%======================================================
%
%
%\begin{figure}[h]
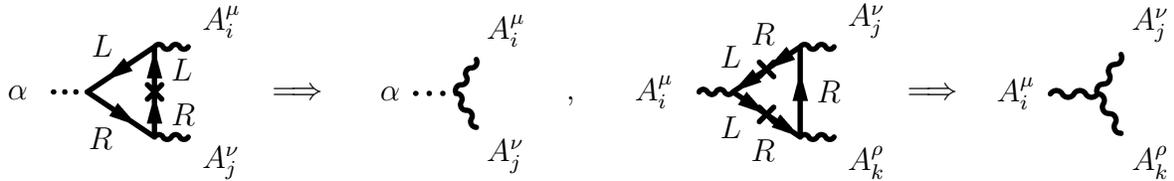
\begin{center}
%\FIGURE[h]{
\begin{tabular}{lcr}
\\
\unitlength=0.3mm
\begin{fmffile}{axionAA_Heavy_2}
\begin{fmfgraph*}(60,40)
\fmfpen{thick} \fmfleft{ii0,ii1,ii2} \fmfstraight \fmffreeze
\fmftop{ii2,t1,t2,t3,oo2} \fmfbottom{ii0,b1,b2,b3,oo1}
\fmf{phantom}{ii2,t1,t2} \fmf{phantom}{ii0,b1,b2}
\fmf{phantom}{t1,v1,b1} \fmf{phantom}{t2,b2} \fmf{phantom}{t3,b3}
\fmffreeze \fmf{dots}{ii1,v1}
\fmf{fermion,label=$L$,l.side=right}{t3,v1}
\fmf{fermion,label=$R$,l.side=right}{b3,a}
\fmf{fermion,label=$L$,l.side=right}{a,t3}
\fmfv{decor.shape=cross,decor.size=.15w}{a}
\fmf{fermion,label=$R$,l.side=right}{v1,b3} \fmf{photon}{b3,oo1}
\fmf{photon}{t3,oo2} \fmffreeze \fmflabel{$\alpha~$}{ii1}
\fmflabel{$A^\mu_i$}{oo2} \fmflabel{$A^\nu_j$}{oo1}
\end{fmfgraph*}
\end{fmffile}
~~~\quad \raisebox{2.8ex}[0cm][0cm]{\unitlength=.65mm
$\Longrightarrow$} \quad ~~
\raisebox{.75ex}[0cm][0cm]{\unitlength=.65mm \unitlength=.3mm
\begin{fmffile}{axionAA_coupling_2}
\begin{fmfgraph*}(30,30)
\fmfpen{thick} \fmfleft{o1} \fmfright{i1,i2}
\fmf{boson}{i1,v1,i2}\fmf{dots}{o1,v1} \fmflabel{$\a$}{o1}
\fmflabel{$A^\nu_j$}{i1} \fmflabel{$A^\mu_i$}{i2}
\end{fmfgraph*}
\end{fmffile}}
\quad ~~ \raisebox{2.8ex}[0cm][0cm]{\unitlength=.65mm ,} \quad
\quad ~~
\unitlength=0.3mm
\begin{fmffile}{GCS_2}
\begin{fmfgraph*}(60,40)
\fmfpen{thick} \fmfleft{ii0,ii1,ii2} \fmfstraight \fmffreeze
\fmftop{ii2,t1,t2,t3,oo2} \fmfbottom{ii0,b1,b2,b3,oo1}
\fmf{phantom}{ii2,t1,t2} \fmf{phantom}{ii0,b1,b2}
\fmf{phantom}{t1,v1,b1} \fmf{phantom}{t2,b2} \fmf{phantom}{t3,b3}
\fmffreeze \fmf{photon}{ii1,v1}
\fmf{fermion,label=$R$,l.side=right}{b3,t3}
\fmf{fermion,label=$R$,l.side=right}{t3,a}
\fmf{fermion,label=$L$,l.side=right}{a,v1}
\fmfv{decor.shape=cross,decor.size=.15w}{a}
\fmf{fermion,label=$L$,l.side=right}{v1,b}
\fmf{fermion,label=$R$,l.side=right}{b,b3}
\fmfv{decor.shape=cross,decor.size=.15w}{b}
%\fmf{fermion,label=$L$,l.side=right}{v1,b3}
\fmf{photon}{b3,oo1} \fmf{photon}{t3,oo2} \fmffreeze
\fmflabel{$A^\mu_i~$}{ii1} \fmflabel{$A^\nu_j$}{oo2}
\fmflabel{$A^\rho_k$}{oo1}
\end{fmfgraph*}
\end{fmffile}%\eea
~~~\quad \raisebox{2.8ex}[0cm][0cm]{\unitlength=.65mm
$\Longrightarrow$} \quad ~~
\raisebox{.3ex}[0cm][0cm]{\unitlength=.65mm \unitlength=.35mm
\begin{fmffile}{GCS_Heavy_2}
\begin{fmfgraph*}(30,30)
\fmfpen{thick} \fmfleft{o1} \fmfright{i1,i2}
\fmf{boson}{i1,v1,i2}\fmf{boson}{o1,v1} \fmflabel{$A^\mu_i$}{o1}
\fmflabel{$A^\rho_k$}{i1} \fmflabel{$A^\nu_j$}{i2}
\end{fmfgraph*}
\end{fmffile}}
\\ \\
\end{tabular}
\end{center}
where we denote by $L$, $R$ the left, right fermions respectively
and by $\times$ the heavy mass insertion.

Consequently, GCS-terms are a prediction of any anomaly-free
chiral gauge theory with light and heavy fermions and it seems
that we cannot distinguish between low-energy predictions of
string theory versus 4d field theory models. However, a deeper
analysis is needed in this direction.

\subsection{Generalized Chern-Simons and the Standard Model}

The presence of GCS terms has important phenomenological
consequences because they provide new anomaly-related couplings
that distinguish these models from others which have been studied
in the literature.

All open string models which attempt to describe the SM contain
various U(1)$'$s, one from each stack of branes that participates
in the construction. A linear combination of all these U(1)$'$s is
the Hypercharge which is anomaly free. However, there are other
linear combinations which are anomalous and their corresponding
gauge boson gain a mass due to the St\"uckelberg mixings with the
axions. As we have argued before, such a configuration needs GCS
terms to cancel all the anomalies.

As an example consider the GCS term $Y\wedge PQ\wedge d PQ$ which
is necessary to cancel the mixed anomalies between the hypercharge
and the Peccei-Quinn anomalous boson. Going from the hypercharge
basis to the photon basis, we perform a rotation that provides
non-trivial couplings of the form $\g Z Z'$ \cite{d-effective}.
Such couplings are of the same order of the $Higgs\to \g\g$
signal, that is the main channel for the discovery of the Higgs.

\section*{Acknowledgements}

%\begin{acknowledgement}
The author would like to thank Massimo Bianchi, Emilian Dudas and
Elias Kiritsis for the fruitful collaboration.
It is a pleasure to thank the organizers of the 2nd Workshop and
Midterm meeting in Napoli "Constituents, Fundamental Forces and
Symmetries of the Universe" 9-13 Oct.2006 and the organizers of
the PRIN meeting in Alessandria 15-16 Dec.2006 for giving the
opportunity to present this work.
This work was supported in part by INFN, by the MIUR-COFIN
contract 2003-023852, by the EU contracts MRTN-CT-2004-503369 and
MRTN-CT-2004-512194, by the INTAS contract 03-516346 and by the
NATO grant PST.CLG.978785.
%\end{acknowledgement}

\end{document}